\documentclass[sigconf,nonacm]{acmart}

\setcopyright{none}
\acmConference[KDD Cup 2026 UniRec Workshop]
{KDD Cup 2026 Tencent UniRec Challenge Workshop}
{August 12, 2026}
{Jeju, Korea}

\acmYear{2026}
\acmDOI{}
\acmISBN{}
\AtBeginDocument{%
  }

\setcopyright{acmlicensed}
\copyrightyear{2018}
\acmYear{2018}
\acmDOI{XXXXXXX.XXXXXXX}
\acmConference[Conference acronym 'XX]{Make sure to enter the correct
  conference title from your rights confirmation email}{June 03--05,
  2018}{Woodstock, NY}
\acmISBN{978-1-4503-XXXX-X/2018/06}

\newcommand\blfootnote[1]{%
  \begingroup
  \renewcommand\thefootnote{}%
  \footnotetext{#1}%
  \addtocounter{footnote}{-1}%
  \endgroup
}

\begin{document}

\title{Topology-Masked Unified Backbone for Joint Feature Interaction and Multi-Domain Sequence Modeling}




\author{Zhihao Zhu}
\authornote{These authors contributed equally to this research.}
\affiliation{%
  \institution{Shandong University}
  \city{Jinan}
  \state{Shandong}
  \country{China}
}
\email{202415670@mail.sdu.edu.cn}

\author{Dezheng Han}
\authornotemark[1]
\affiliation{%
  \institution{Shandong University}
  \city{Jinan}
  \state{Shandong}
  \country{China}}
\email{dezhenghan@mail.sdu.edu.cn}

\author{Jikang Xia}
\authornotemark[1]
\affiliation{%
  \institution{Tsinghua University}
  \city{Beijing}
  \country{China}}
\email{xiajk24@mails.tsinghua.edu.cn}

\author{Shuaishuai Guo}
\correspondingauthor
\affiliation{%
  \institution{Shandong University}
  \city{Jinan}
  \state{Shandong}
  \country{China}
}
\email{shuaishuai\_guo@sdu.edu.cn}

\renewcommand{\shortauthors}{Zhu, Han, and Xia}

\begin{abstract}
Large-scale post-click conversion rate (CVR) prediction requires jointly modeling heterogeneous feature interactions and dependencies over multi-domain user behavior sequences. Existing industrial ranking models usually handle these two aspects with separate modules. Recent unified architectures attempt to incorporate them into a single framework, but such unification often relies on coordination between modules and does not fully organize all information sources within the same interaction space. 
To address this problem, we propose MaskRec, a topology-masked unified token interaction architecture for feature interaction and multi-domain sequence modeling. MaskRec transforms heterogeneous features, multi-domain behavior sequences, and contextual signals into unified token representations, and further introduces learnable global memory tokens and domain-level memory tokens as information aggregation nodes. Based on this unified token space, MaskRec designs a structured attention mask, TopoMask, which selectively enables or blocks attention connections according to the structural differences and modeling requirements of different information sources. In this way, heterogeneous feature interaction and multi-domain sequence modeling are performed within the same topology-constrained attention process.
In addition, MaskRec incorporates a dual-path interactive query generation module to inject candidate-conditioned user--item interaction signals before the unified backbone. Experiments on the Tencent Advertising Algorithm Competition dataset show that MaskRec achieves stable improvements over the official baseline, validating the effectiveness of the proposed unified framework for industrial CVR prediction.
\end{abstract}





\maketitle
\blfootnote{\scriptsize
KDD Cup 2026 Tencent UniRec Challenge Workshop, August 12, 2026, Jeju, Korea.\\
Competition website: \url{https://algo.qq.com/}.\\
Our code is publicly available at \url{https://github.com/zzhlkw-ai/TAAC2026}.
}

\section{Introduction}
Post-click conversion rate (CVR) prediction is a core task in industrial advertising systems \cite{ma2018esmm}. Its goal is to estimate the probability that a user will further convert after clicking a candidate advertisement or item. Since the input information is highly heterogeneous in semantic granularity, statistical distribution, and temporal structure, early industrial ranking models typically handle non-sequential feature interaction and behavior sequence modeling with separate modules. Although this separated design is efficient and stable in engineering practice, it also limits deep interactions among different information sources \cite{huang2026hyformer,zhang2026onetrans}. Therefore, recent industrial recommendation studies have begun to explore more unified modeling frameworks that jointly handle feature interaction and sequence modeling within a shared architecture \cite{huang2026mixformer,wang2026lens}.


However, existing unified architectures still do not sufficiently address how heterogeneous information should be organized in a unified space. Although related methods enhance the interaction between sequence representations and non-sequential features through unified backbones or hybrid structures \cite{huang2026hyformer,zhang2026onetrans}, such unification either overlooks the structural differences and interaction boundaries among heterogeneous information in a unified sequence, or relies on alternating coordination between modeling processes, without truly realizing unified modeling within the same space. For large-scale CVR prediction, a more desirable unified architecture should not only break the modular boundary between feature interaction and sequence modeling, but also preserve the structural differences of different information sources and organize interactions among them during unified modeling.


To address the above issues, we propose MaskRec, a topology-masked unified token interaction architecture for feature interaction and multi-domain sequence modeling. MaskRec first transforms heterogeneous features, multi-domain behavior sequences, and contextual signals into unified token representations, and further introduces learnable global memory tokens and domain-level memory tokens as information aggregation nodes. Unlike unconstrained fully connected self-attention over all tokens, MaskRec designs a structured attention mask, TopoMask, whose core idea is inspired by causal attention \cite{vaswani2017attention}. Specifically, TopoMask selectively enables or blocks attention connections according to the structural differences and modeling requirements of different information sources. As a result, MaskRec can preserve the internal structure of different information sources while enabling controlled cross-source interactions within a unified token space.


In addition to the topology-masked unified backbone, MaskRec introduces a dual-path interactive query generation module. It should be noted that the query tokens in this paper are not the query vectors used in attention computation, but interaction-enhanced tokens constructed before entering the unified backbone. They are used to inject conditional interaction information between users and candidate items, and are further transformed into enhanced query tokens for different behavior domains.


We evaluate MaskRec on the Tencent Advertising Algorithm Competition dataset \cite{tencent2026competition}. Experimental results show that MaskRec achieves stable performance improvements over the official baseline, indicating that topology-masked token routing can effectively unify feature interaction and multi-domain sequence modeling in industrial CVR prediction.


Our main contributions are summarized as follows:


\begin{itemize}
    \item We propose MaskRec, a topology-masked unified token interaction architecture that incorporates heterogeneous features, multi-domain behavior sequences, contextual signals, query tokens, and memory tokens into a shared interaction space for unified modeling.


    \item We design TopoMask, a structured attention mask that explicitly defines token-level query-key connections according to token roles and behavior-domain relationships, enabling intra-source structural modeling, domain-level aggregation, and memory-mediated cross-token interaction within the same attention process.


    \item We propose a dual-path query generation module that constructs base queries containing user--item interaction information through bidirectional cross-attention between user-side and candidate-item-side tokens, and further generates domain-aware queries through FiLM modulation conditioned on behavior domains.


    \item Experiments on the Tencent Advertising Algorithm Competition dataset show that MaskRec achieves stable performance improvements over the official baseline, validating the effectiveness of topology-constrained token routing in large-scale industrial CVR prediction.

\end{itemize}

The rest of this paper is organized as follows. Section~2 reviews related work. Section~3 presents the design of MaskRec. Section~4 reports the experimental results. Section~5 concludes the paper and discusses future directions.


\section{Related Work}

\subsection{Sequence Modeling and Feature Interaction in Industrial Ranking}
Industrial ranking models usually combine two types of modeling components: behavior sequence modeling and non-sequential feature interaction. Candidate-aware models such as DIN adaptively aggregate user behaviors with respect to the target item \cite{zhou2018deep}, while self-attentive sequential models such as SASRec and long-sequence models further improve the capacity of user behavior modeling \cite{kang2018sasrec}. These methods are effective for capturing user interests from historical behaviors, especially when the relevance between the candidate item and past behaviors is critical for prediction. In parallel, feature interaction models such as DCNv2, RankMixer, and token-mixing architectures focus on modeling high-order interactions among heterogeneous non-sequential features, including user attributes, item attributes, context signals, and statistical features \cite{wang2021dcnv2,zhu2025rankmixer}. These models are widely used in CTR/CVR prediction and industrial recommendation systems, as they provide efficient and scalable mechanisms for modeling feature crossing patterns.

However, existing industrial ranking models usually handle sequence modeling and feature interaction with relatively separate modules: behavior sequences are first encoded or compressed, and then fused with non-sequential features in downstream interaction layers. Although this design is efficient and stable, it weakens early and deep interactions between sequential and non-sequential information \cite{huang2026hyformer,zhang2026onetrans}.

\subsection{Unified Recommendation Architectures}
Recent studies have begun to unify sequence modeling and feature interaction within a shared architecture. HyFormer integrates long-sequence modeling and feature interaction through an alternating process of query decoding and query boosting, thereby alleviating the late-fusion limitation of traditional pipelines \cite{huang2026hyformer}. OneTrans further tokenizes sequential and non-sequential attributes into a unified token stream and models them with a single Transformer backbone \cite{zhang2026onetrans}. MixFormer and LENS also reflect the same trend of organizing recommendation models around latent-query or unified-interaction paradigms \cite{huang2026mixformer,wang2026lens}. These methods represent important steps toward unified industrial ranking models.

Nevertheless, existing unified architectures still differ in how they organize heterogeneous information in the unified space. Hybrid architectures mainly achieve unification through the coordination of different modeling processes, while unified-backbone architectures often rely on relatively general token interaction mechanisms once heterogeneous inputs are placed into the same token sequence. As a result, the structural differences and interaction boundaries among heterogeneous information sources are not sufficiently characterized. In contrast, MaskRec focuses on the organization of heterogeneous tokens within the unified space, and introduces TopoMask to explicitly regulate token-level information flow according to structural modeling requirements.

\begin{figure*}[t]
    \centering
    \includegraphics[width=\textwidth]{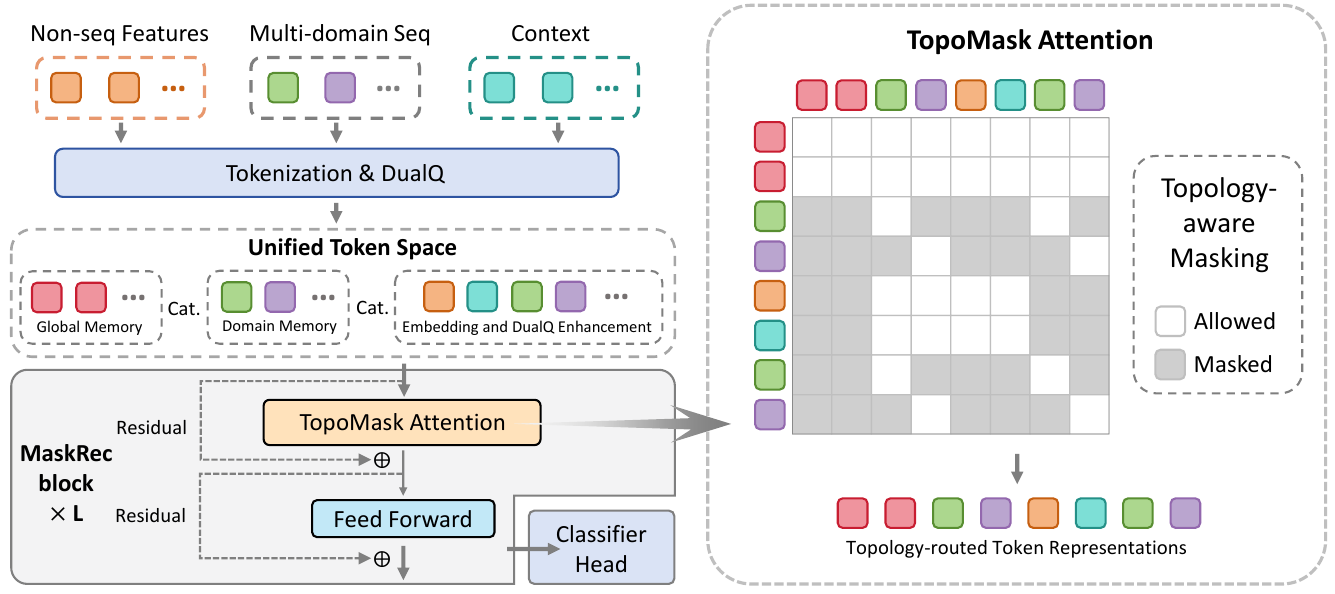}
    \caption{Overall architecture of MaskRec. Heterogeneous non-sequential features, multi-domain behavior sequences, and contextual signals are first transformed into a unified token space together with memory and query-enhancement tokens. The unified token sequence is then processed by stacked MaskRec blocks, where TopoMask Attention selectively enables or blocks token-level attention connections according to the structural differences and modeling requirements of different information sources. This topology-aware routing allows MaskRec to preserve source-specific structures while enabling controlled cross-source interactions for industrial CVR prediction.}
    \label{fig:maskrec_overview}
\end{figure*}

\section{Method}

\subsection{Problem Formulation}

We consider post-click conversion rate (CVR) prediction in a large-scale industrial advertising system. For each clicked impression, the model receives heterogeneous non-sequential features, contextual signals, and multiple behavior sequences from different domains. Let $\mathcal{X}_{ns}$ denote the non-sequential features, including user-side attributes, candidate-item-side attributes, and their associated context features. Let $\mathcal{S}=\{S_1,S_2,\ldots,S_K\}$ denote $K$ behavior domains, where each sequence $S_k=\{s_{k,1},s_{k,2},\ldots,s_{k,L_k}\}$ contains user historical behaviors from the $k$-th domain. Given these inputs, the goal is to estimate the probability that the user will convert after clicking the candidate item:
\begin{equation}
    \hat{y}=f_{\theta}(\mathcal{X}_{ns}, \mathcal{S}),
\end{equation}
where $\hat{y}\in[0,1]$ is the predicted conversion probability and $\theta$ denotes the model parameters.

The key challenge lies in how to jointly model heterogeneous feature interactions and multi-domain behavior dependencies in a unified yet structured manner. Ideally, a unified ranking backbone should place all information sources into a shared token space,
\begin{equation}
    T = [T_{ns}; T_c; T_1; \ldots; T_K],
\end{equation}
where $T_{ns}$ denotes non-sequential feature tokens, $T_c$ denotes contextual tokens, and $T_k$ denotes behavior tokens from the $k$-th domain. The backbone then updates token representations through a shared interaction operator:
\begin{equation}
    T^{(l+1)}=\mathcal{B}(T^{(l)}), \quad l=0,\ldots,L-1.
\end{equation}
However, $\mathcal{B}$ should not be an unconstrained full-interaction operator over all tokens. Since different information sources have distinct semantic roles, statistical properties, and sequential structures, the desired operator should simultaneously preserve source-specific structures and support controlled cross-source interactions. This requirement motivates the design of a topology-aware token interaction mechanism, where heterogeneous signals are modeled in the same token space while their information flow is explicitly regulated.

\begin{figure*}[t]
    \centering
    \includegraphics[width=\textwidth]{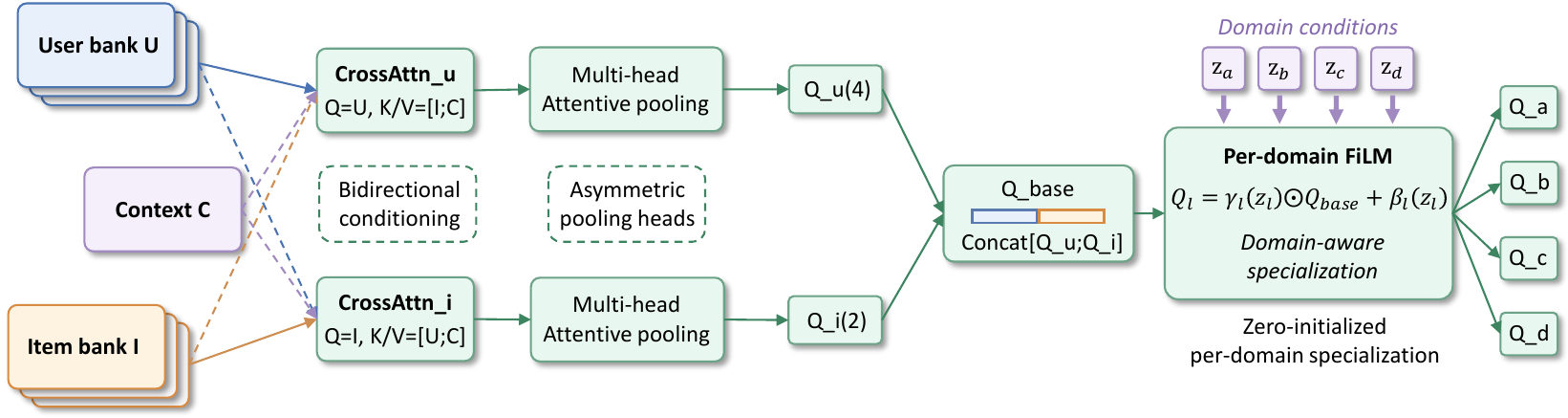}
    \caption{Dual-path interactive query generation module DualQ. User-side, candidate-item-side, and contextual non-sequential tokens are first conditionally modeled through bidirectional cross-attention. The resulting representations are summarized by asymmetric attentive pooling heads to form a base interaction query, which is further specialized with domain conditions to produce domain-aware query tokens. These query tokens are interaction-enhanced tokens constructed before the unified backbone, rather than the query vectors used inside attention computation.}
    \label{fig:dualq_generator}
\end{figure*}

\subsection{Overall Architecture}

Figure~\ref{fig:maskrec_overview} illustrates the overall architecture of MaskRec. Given a clicked impression, the input consists of heterogeneous non-sequential features, contextual signals, and multi-domain behavior sequences. MaskRec first converts these inputs into token representations in a shared embedding dimension. Non-sequential features are embedded as feature tokens, contextual signals are represented as context tokens, and each behavior domain is transformed into a sequence of domain-specific behavior tokens. In parallel, the dual-path interactive query generation module constructs enhanced query tokens from user-side, candidate-item-side, and contextual information, providing candidate-conditioned auxiliary tokens for subsequent multi-domain sequence modeling.

After tokenization, MaskRec organizes all tokens into a unified token space. In addition to input-derived tokens, MaskRec introduces two types of learnable memory tokens. Global memory tokens are shared across all information sources and serve as global aggregation nodes for collecting information from the entire token space. Domain-level memory tokens are associated with individual behavior domains and are used to summarize domain-specific sequence information. The full token sequence is therefore composed of global memory tokens, domain-level memory tokens, non-sequential feature tokens, context tokens, enhanced query tokens, and multi-domain behavior tokens.

The unified token sequence is then fed into a stack of MaskRec blocks. Each block contains a TopoMask Attention layer followed by a feed-forward network with residual connections and normalization. During attention computation, TopoMask selectively enables or blocks token-level query-key connections according to the structural differences and modeling requirements of different information sources. This allows source-local modeling, domain-level aggregation, and controlled cross-source interaction to be performed within the same backbone, rather than relying on separated sequence modeling and feature interaction modules.

After the final MaskRec block, the output token representations are aggregated for prediction. MaskRec mainly uses the representations of global memory tokens, domain-level memory tokens, and enhanced query tokens as the readout signals, since they respectively capture global context, domain-specific sequence summaries, and candidate-conditioned user--item interaction information. These representations are concatenated and passed into a classifier head to produce the final CVR prediction.

\subsection{Dual-Path Interactive Query Generation}

To provide candidate-conditioned interaction signals before unified token fusion, MaskRec introduces a dual-path interactive query generation module, DualQ. As illustrated in Figure~\ref{fig:dualq_generator}, let $U\in\mathbb{R}^{n_u\times d}$, $I\in\mathbb{R}^{n_i\times d}$, and $C\in\mathbb{R}^{n_c\times d}$ denote the user-side, candidate-item-side, and contextual non-sequential token banks, respectively. DualQ performs bidirectional conditional modeling between the user side and the candidate-item side through cross-attention:
\begin{equation}
    H_u=\mathrm{CrossAttn}(U,[I;C],[I;C]),
\end{equation}
\begin{equation}
    H_i=\mathrm{CrossAttn}(I,[U;C],[U;C]),
\end{equation}
where $[\cdot;\cdot]$ denotes token concatenation. This bidirectional design allows user-side tokens to be conditioned on the candidate item and context, while candidate-item-side tokens can also receive user-side contextual information.

The resulting representations are then summarized by attentive pooling heads to obtain compact interaction tokens:
\begin{equation}
    Q_u=\mathrm{Pool}_u(H_u), \quad Q_i=\mathrm{Pool}_i(H_i),
\end{equation}
where $Q_u\in\mathbb{R}^{m_u\times d}$ and $Q_i\in\mathbb{R}^{m_i\times d}$ denote user-side and item-side interaction queries. The base interaction query bank is formed as:
\begin{equation}
    Q_{\mathrm{base}}=[Q_u;Q_i].
\end{equation}

Since different behavior domains may require different interaction signals, MaskRec further specializes the base queries with domain conditions. For the $k$-th behavior domain, a domain condition vector $z_k$ is used to modulate $Q_{\mathrm{base}}$ through feature-wise linear modulation:
\begin{equation}
    Q_k=\gamma_k(z_k)\odot Q_{\mathrm{base}}+\beta_k(z_k),
\end{equation}
where $\gamma_k(\cdot)$ and $\beta_k(\cdot)$ are learnable modulation functions, and $\odot$ denotes element-wise multiplication. The generated $Q_k$ serves as the domain-aware enhanced query tokens for the $k$-th domain. These tokens are then inserted into the unified token space together with non-sequential tokens, sequence tokens, and memory tokens.

DualQ plays the role of an interaction-aware query initializer for the unified backbone. Instead of generating query tokens from user or item features independently, it explicitly conditions the two sides on each other before token-level fusion. Therefore, the generated query tokens already encode candidate-specific user--item interaction signals, while the subsequent domain-wise modulation allows them to adapt to behavior domains with different semantics. This design makes the unified backbone receive not only raw heterogeneous tokens, but also compact auxiliary tokens that summarize candidate-conditioned interaction evidence for multi-domain sequence modeling.

\subsection{Unified Token Space}
MaskRec represents all input information as tokens in a shared interaction space. Specifically, the non-sequential features are embedded and grouped into non-sequential tokens:
\begin{equation}
    T_{ns}=\mathrm{Embed}(\mathcal{X}_{ns}).
\end{equation}
For each behavior domain $k$, the behavior sequence $S_k$ is encoded into a sequence token set:
\begin{equation}
    T_k=\mathrm{SeqEmbed}_k(S_k),
\end{equation}
where $T_k\in\mathbb{R}^{L_k\times d}$ contains the token representations of behaviors in the $k$-th domain.

To support information aggregation and cross-source communication, MaskRec introduces learnable memory tokens. Global memory tokens $M_g\in\mathbb{R}^{m_g\times d}$ are used to aggregate information from the entire token space, while domain-level memory tokens $M_k\in\mathbb{R}^{m_d\times d}$ are used to summarize domain-specific sequence information. The full token sequence is then constructed as:
\begin{equation}
    T=[M_g;M_1;\ldots;M_K;T_{ns};Q_1;\ldots;Q_K;T_1;\ldots;T_K].
\end{equation}
This unified token construction brings heterogeneous inputs into a common interaction space, allowing non-sequential features, contextual signals, enhanced query tokens, memory tokens, and multi-domain behavior sequences to be optimized within the same backbone. Such a design avoids the rigid separation between feature interaction and sequence modeling, and makes it possible for different information sources to exchange useful signals across layers. However, simply placing all tokens into one sequence is not sufficient. Since these tokens differ in semantic roles, structural properties, and modeling requirements, fully connected self-attention may introduce noisy or inappropriate interactions. Therefore, the unified token space further requires a structured routing mechanism to regulate token-level information flow, which motivates the design of TopoMask Attention.

\subsection{TopoMask Attention}

TopoMask Attention is the core component of MaskRec. It generalizes masked attention from temporal order constraints to topology-aware interaction constraints. Given the input token matrix $T\in\mathbb{R}^{N\times d}$, standard self-attention computes:
\begin{equation}
    \mathrm{Attn}(T)=\mathrm{Softmax}\left(\frac{QK^\top}{\sqrt{d}}\right)V,
\end{equation}
where $Q=TW_Q$, $K=TW_K$, and $V=TW_V$. This formulation allows every token to attend to every other token, which may not be suitable for heterogeneous industrial ranking inputs.

TopoMask introduces a structural mask matrix $A\in\{0,1\}^{N\times N}$ to define whether a query token is allowed to attend to a key token. The corresponding additive attention mask is defined as:
\begin{equation}
    \mathcal{M}_{ij}=
    \begin{cases}
    0, & A_{ij}=1,\\
    -\infty, & A_{ij}=0.
    \end{cases}
\end{equation}
The topology-masked attention is then computed as:
\begin{equation}
    \mathrm{TopoAttn}(T)=\mathrm{Softmax}\left(\frac{QK^\top}{\sqrt{d}}+\mathcal{M}\right)V.
\end{equation}
For the heterogeneous inputs considered in this task, we configure the TopoMask connectivity according to their structural differences and interaction requirements. Specifically, the adopted topology follows three design considerations. First, tokens from the same information source or behavior domain should preserve sufficient local interactions to model their internal structure. Second, domain-level aggregation nodes should collect information from their corresponding behavior domains and interact with domain-aware query tokens. Third, cross-source interactions should follow controlled paths rather than unrestricted full attention, so that heterogeneous sources can exchange useful information without introducing excessive noisy interactions.

Specifically, global memory tokens are allowed to attend to the entire token space for global information aggregation. Non-sequential and contextual feature tokens retain interactions within the feature-side group. For each behavior domain, the corresponding domain memory, enhanced query tokens, and behavior tokens form a domain-specific interaction group and are jointly modeled. Direct interactions between unrelated behavior domains are blocked, while candidate-conditioned cross-source evidence is introduced through DualQ and broader information is aggregated by the global memory tokens. This connectivity pattern represents one concrete strategy under TopoMask for the heterogeneous input structure of the current task and can be modified for other settings by redefining the allowed query-key connections.

With this design, TopoMask Attention enables source-local modeling, domain-aware aggregation, and controlled cross-source interaction within the same attention layer. Compared with separated sequence modeling and feature interaction modules, TopoMask allows heterogeneous signals to be jointly optimized in a unified backbone. Compared with unconstrained unified token streams, it explicitly regulates information flow and better preserves the structural differences among heterogeneous inputs.

\subsection{MaskRec Block and Prediction}

A MaskRec block applies TopoMask Attention followed by a feed-forward network:
\begin{equation}
    \tilde{T}^{(l)} = T^{(l)} + \mathrm{TopoAttn}(\mathrm{Norm}(T^{(l)})),
\end{equation}
\begin{equation}
    T^{(l+1)} = \tilde{T}^{(l)} + \mathrm{FFN}(\mathrm{Norm}(\tilde{T}^{(l)})),
\end{equation}
where $l$ denotes the layer index. By stacking multiple MaskRec blocks, the model progressively refines token representations under topology-constrained information routing.

After the final MaskRec block, MaskRec aggregates the output representations from selected tokens, including global memory tokens, domain memory tokens, and interaction-enhanced query tokens. The aggregated representation is fed into a prediction head to estimate the conversion probability:
\begin{equation}
    h=\mathrm{Aggregate}(T^{(L)}),
\end{equation}
\begin{equation}
    \hat{y}=\sigma(\mathrm{MLP}(h)),
\end{equation}
where $\sigma(\cdot)$ denotes the sigmoid function.

The model is trained with the binary cross-entropy loss:
\begin{equation}
    \mathcal{L}
    =-\frac{1}{|\mathcal{D}|}\sum_{(x,y)\in\mathcal{D}}
    \left[y\log \hat{y}+(1-y)\log(1-\hat{y})\right],
\end{equation}
where $\mathcal{D}$ denotes the training set and $y\in\{0,1\}$ is the ground-truth conversion label.

\section{Experiments}
\subsection{Dataset and Evaluation Metric}

We evaluate MaskRec on the dataset released for the Tencent Advertising Algorithm Competition. The task is post-click conversion rate prediction on large-scale anonymized advertising logs, where the model estimates the probability that a user performs a conversion action after clicking a candidate advertisement or item. Following the partition strategy of the official baseline, we split the released second-stage training data into training and validation subsets with a ratio of $0.9{:}0.1$. We report validation AUC on this held-out subset and test AUC returned by the official evaluation platform.

The area under the ROC curve (AUC) is used as the primary evaluation metric. Given a positive sample set $P$ and a negative sample set $N$, AUC can be written as
\begin{equation}
    \mathrm{AUC}
    =
    \frac{1}{|P||N|}
    \sum_{i\in P}
    \sum_{j\in N}
    \mathbf{1}
    \left[
    \hat{p}_i > \hat{p}_j
    \right],
\end{equation}
where $\hat{p}_i$ and $\hat{p}_j$ denote the predicted scores of positive and negative samples, respectively. AUC measures the probability that a randomly selected positive sample is ranked above a randomly selected negative sample, making it suitable for the imbalanced and ranking-oriented CVR prediction setting.

Since training and test evaluation are conducted on the official competition platform, we focus on a compact set of controlled comparisons covering the overall architecture, feature-side modeling choices, temporal information, and memory-token capacity. The results are summarized in Table~\ref{tab:main_results}.





\begin{table}[t]
\centering
\caption{Overall comparison, ablation results, and sensitivity to memory-token capacity on the competition dataset.}
\label{tab:main_results}
\small
\begin{tabular}{lcc}
\toprule
Variant & Val AUC & Test AUC \\
\midrule
\multicolumn{3}{l}{\textbf{(A) Overall comparison}} \\
HyFormer-style baseline & 0.831827 & 0.824902 \\
MaskRec & \textbf{0.841253} & \textbf{0.834640} \\
\midrule
\multicolumn{3}{l}{\textbf{(B) Ablation study}} \\
Full model & \textbf{0.841253} & \textbf{0.834640} \\
w/o DualQ & 0.840037 & 0.833665 \\
Dense features $\rightarrow$ one token & 0.831861 & 0.825470 \\
w/o selected user/item ID features & 0.839800 & 0.833275 \\
w/o temporal features & 0.841045 & 0.834468 \\
\midrule
\multicolumn{3}{l}{\textbf{(C) Memory token sensitivity}} \\
32 / 8 & \textbf{0.841253} & \textbf{0.834640} \\
16 / 4 & 0.841090 & 0.834589 \\
8 / 2 & 0.840914 & 0.834324 \\
\bottomrule
\end{tabular}
\end{table}

\subsection{Overall Comparison}
We compare MaskRec with a HyFormer-style baseline that retains the same feature-processing and input-construction pipeline while replacing the proposed unified backbone with the original HyFormer-style architecture. As shown in Table~\ref{tab:main_results}(A), MaskRec improves validation AUC from 0.831827 to 0.841253 and test AUC from 0.824902 to 0.834640. This result demonstrates the advantage of the proposed unified interaction architecture for organizing heterogeneous features and multi-domain behavior signals.



\subsection{Ablation Study}

We next evaluate several feature-side and temporal components used in the competition system. Since the experiments were conducted on the official platform, we report a compact set of controlled variants rather than an exhaustive combinatorial ablation. The results are summarized in Table~\ref{tab:main_results}(B).

\paragraph{Dual-path query generation.}
We remove DualQ and revert to the original query generation mechanism. This variant obtains 0.840037 validation AUC and 0.833665 test AUC, corresponding to decreases of 0.001216 and 0.000975, respectively. The consistent degradation indicates that candidate-conditioned user--item interaction signals provide effective auxiliary information before unified token modeling.

\paragraph{Dense feature organization.}
We collapse the dense-side representation into a single token, removing grouped dense tokens, structured dense representations, aligned pair features, and the fid-87 residual path. This variant reaches 0.831861 validation AUC and 0.825470 test AUC, substantially below the full configuration. The result suggests that heterogeneous dense features benefit from preserving their internal organization rather than being compressed into a single homogeneous representation.

\paragraph{Sparse integer features.}
We remove a selected subset of user-side and item-side integer fields. The resulting model obtains 0.839800 validation AUC and 0.833275 test AUC. The consistent degradation indicates that sparse identifier features provide complementary information beyond dense representations and behavior sequences.

\paragraph{Temporal features.}
We remove the temporal feature family, including global-time signals and sequence time-gap related features. This variant reaches 0.841045 validation AUC and 0.834468 test AUC. Although the performance drop is modest, the consistent degradation on both splits suggests that temporal context provides complementary information for modeling time-dependent user behavior and conversion patterns.

\subsection{Sensitivity to Memory Token Capacity}
MaskRec introduces global memory tokens and domain-level memory tokens as learnable aggregation nodes. We study memory-token capacity by progressively reducing the numbers of global and domain-level memory tokens. Starting from the default configuration of $32/8$, we reduce the numbers to $16/4$ and $8/2$, where the two values denote the numbers of global and domain-level memory tokens, respectively. The results are reported in Table~\ref{tab:main_results}(C).

The performance decreases gradually as memory tokens are reduced. The $16/4$ configuration remains close to the default, while further reducing the capacity to $8/2$ produces clearer degradation, particularly on the test set. This trend suggests that memory tokens provide useful aggregation capacity for heterogeneous information and multi-domain behaviors. At the same time, the relatively small gap between $32/8$ and $16/4$ indicates that MaskRec is not overly sensitive to moderate reductions in memory capacity.



\balance

\section{Conclusion and Discussion}
In this paper, we proposed MaskRec, a topology-masked unified token interaction architecture for large-scale post-click conversion rate prediction. MaskRec organizes heterogeneous features, contextual signals, multi-domain behavior sequences, memory tokens, and enhanced query tokens into a shared token space, and introduces TopoMask to regulate token-level information flow with structured attention masks. Together with the DualQ module for candidate-conditioned query generation, MaskRec enables unified yet controlled modeling of feature interaction and multi-domain sequence dependencies. Experiments on the Tencent Advertising Algorithm Competition dataset show that MaskRec achieves stable improvements over the official baseline, demonstrating the effectiveness of topology-constrained token routing.
One limitation of MaskRec is that the topology in TopoMask is manually specified according to empirical structural priors of industrial recommendation data. Although this design is simple and interpretable, it may not be optimal across different datasets or feature systems. Future work will explore learnable or adaptive topology construction, allowing token routing patterns to be adjusted automatically from data.

\section*{Acknowledgments}

We thank the organizers of the Tencent Advertising Algorithm Competition for providing the large-scale anonymized advertising dataset, the official baseline system, and the evaluation platform. We also appreciate their efforts in organizing a practical benchmark for studying unified sequence modeling and feature interaction in industrial recommendation scenarios.

\bibliographystyle{ACM-Reference-Format}
\bibliography{references}

\clearpage
\appendix
\setcounter{section}{0}
\setcounter{subsection}{0}
\setcounter{subsubsection}{0}
\setcounter{figure}{0}
\setcounter{table}{0}
\setcounter{equation}{0}

\renewcommand{\thefigure}{A.\arabic{figure}}
\renewcommand{\thetable}{A.\arabic{table}}
\renewcommand{\theequation}{A.\arabic{equation}}

\section{Appendix}


This appendix reports supplementary materials from our Tencent Advertising Algorithm Competition system. During the competition, our exploration was organized as a set of practical modeling ideas built on the official HyFormer-style baseline, including dual-query generation, field-aware dense tokenization, temporal context modeling, long-tail frequency bucketing, and optimization stabilization. We further document implementation-level ideas and fine-grained design choices that were explored during system development but are not detailed in the main text. The following sections present the corresponding dataset analysis, implementation details, experimental settings, and competition-stage results.

\subsection{Dataset Details}
\label{app:dataset}

\subsubsection{Competition Task}

This work is evaluated on the Tencent Advertising Algorithm Competition, an industrial post-click conversion ranking benchmark released on anonymized advertising logs \cite{tencent2026competition}. For each clicked ad exposure, the task is to estimate the probability that the user subsequently performs a valuable post-click action, such as a purchase, form submission, or app installation, on the advertised item. Submissions are ranked by ROC-AUC on a held-out test set.

The task has two important characteristics. First, the conversion label is observed only after a click, so the prediction setting inherits the sample-selection bias of post-click conversion estimation \cite{ma2018esmm}. Second, the released data is strongly class-imbalanced: only about eight percent of clicked exposures are positive conversion samples. This makes ranking-oriented metrics such as AUC more informative than accuracy and motivates careful optimization under class imbalance.

\subsubsection{Dataset Scale, Labels, and Feature Schema}

The released dataset contains an industrial-scale set of fully anonymized advertising samples. The training split contains approximately $1.978\times10^{7}$ rows, and the validation split contains approximately $2.214\times10^{6}$ rows. Both splits span the same six-day window from 2026-03-27 to 2026-04-01 in Beijing time, with second-level timestamps. The validation split is therefore a same-window row-group split rather than a future-period split, so the reported validation AUC mainly reflects in-distribution generalization rather than temporal extrapolation.

The positive conversion rate is approximately $8.06\%$ in both splits, meaning that negative samples outnumber positive samples by about $11.4\times$. Each row contains sparse integer fields for the user and the candidate item, dense side-information fields for both sides, pairwise integer and dense fields, and multi-domain behavioral sequences with timestamps. The item-dense bundle contains 770 dimensions and includes several schema-defined dense fields. The schema is parsed from a sidecar \texttt{schema.json} and is shared across training, validation, and inference.

\subsubsection{Temporal and Behavioral Patterns}

Although the dataset spans only six days, temporal structure is predictive at multiple granularities. Traffic volume is highly uneven across the day. Early morning hours contribute only a small fraction of samples, while evening hours are the busiest. However, the conversion-rate peak is not aligned with the traffic peak: conversions are more concentrated in late-morning periods, whereas the evening traffic peak shows a lower conversion rate. This mismatch between traffic volume and conversion probability suggests that absolute and cyclic time signals carry label-correlated information beyond relative sequence positions.

Behavior histories are organized into multiple domains with different discriminative strengths. Table~\ref{tab:app_seqlen} reports the average sequence length per domain split by label. Domains a and b show clear length-based separation between negative and positive samples, while domains c and d show weak or nearly no separation. This asymmetry motivates domain-aware modeling: different behavior domains should not necessarily share identical temporal, aggregation, or interaction patterns.

\subsection{Supplementary Component Details}
\label{app:components}

This section summarizes the main implementation ideas explored during the competition. Some components were retained in the final competition configuration, while others were evaluated as targeted architectural variants. Together, they cover feature tokenization, query construction, sequence modeling, temporal context, long-tail representation, and optimization.

\subsubsection{Competition-Stage Initialization and Preliminary-Stage Transfer}
\label{app:transfer}

The second-stage system was initialized by transferring a set of modeling choices that had already been explored during the preliminary stage, while adapting the training configuration to the larger multi-GPU setting. We also corrected the item-dense loading path so that schema-defined item-dense features were actually populated from the released data rather than bypassed by an empty schema path.

For the seven-GPU configuration, the per-GPU batch size was adjusted to maintain a suitable effective global batch size, the learning rate was rescaled accordingly, the number of epochs was shortened to keep the overall optimization budget comparable to the second-stage baseline, and the EMA start step was aligned with warmup. This configuration serves as the starting point for the subsequent competition-stage experiments.

\subsubsection{ID-Frequency Bucketing}
\label{app:idfreq}

Industrial advertising data contains long-tail identifiers, where many user and item IDs appear only a few times. To provide frequency-aware side information, each identifier is assigned to a discrete bucket according to its occurrence count in the training data.

Let $c_v$ denote the occurrence count of identifier $v$. Given bucket boundaries $\tau_1<\cdots<\tau_{K-1}$, the bucket index is defined as
\begin{equation}
    \mathrm{buck}(v)
    =
    1+\sum_{k=1}^{K-1}\mathbf{1}[c_v>\tau_k],
\end{equation}
with bucket 0 reserved for padding and unseen identifiers. A shared frequency-bucket embedding table then supplies an additional representation for identifiers with similar occurrence frequencies. This allows rare IDs to share statistical strength without removing their original identity embeddings.

\subsubsection{Capacity Scaling}
\label{app:capacity}

We also explored moderate scaling of the interaction backbone. In the corresponding configuration, the model dimension was increased from 128 to 192, the number of query tokens from 4 to 6, and the number of user-side non-sequential tokens from 11 to 14. The purpose of this experiment was to increase both hidden representation capacity and the number of compact interaction units available to the downstream fusion process.

\subsubsection{Non-Sequential and Field-Wise Dense Tokenization}
\label{app:ns_tokenization}

The competition baseline uses RankMixer-style non-sequential tokenization to map heterogeneous non-sequential features into a compact token bank. Rather than directly applying quadratic self-attention to all raw feature fields, the tokenizer projects and mixes the feature representations into a fixed number of user-side and item-side tokens.

For item-dense inputs, we further explored a field-wise tokenizer. Instead of projecting the entire 770-dimensional item-dense bundle through a single transformation, each schema-defined dense field is independently processed by a field-specific normalization and projection path:
\begin{equation}
    \mathbf{t}_f
    =
    \operatorname{LN}_d
    \left(
    \operatorname{SiLU}
    \left(
    W_f \operatorname{LN}(\mathbf{x}_f)+\mathbf{b}_f
    \right)
    \right).
\end{equation}
This design preserves field-level distinctions before downstream interaction and follows the broader field-as-token perspective of tabular representation learning \cite{gorishniy2021revisiting}.

\begin{table}[t]
\centering
\caption{Average behavioral sequence length per domain, split by label. Domains a and b show stronger class separation, while domains c and d are less discriminative by length alone.}
\label{tab:app_seqlen}
\small
\begin{tabular}{lccc}
\toprule
Domain & Neg. avg & Pos. avg & Separation \\
\midrule
a & 506 & 598 & strong \\
b & 324 & 368 & moderate \\
c & 395 & 397 & weak \\
d & 460 & 458 & weak \\
\bottomrule
\end{tabular}
\end{table}

\subsubsection{Dual-Query Generation with Cross-Side Conditioning}
\label{app:dualq}

During the competition, we explored a dual-query generation module, DualQ, which constructs query tokens from separate user-side and item-side branches. Let $\mathbf{U}\in\mathbb{R}^{n_u\times d}$, $\mathbf{I}\in\mathbb{R}^{n_i\times d}$, and $\mathbf{C}\in\mathbb{R}^{n_c\times d}$ denote the user-side, item-side, and contextual token banks. DualQ first performs bidirectional cross-side conditioning:
\begin{align}
    \tilde{\mathbf{U}} &= \operatorname{CrossAttn}_u(\mathbf{U}, [\mathbf{I};\mathbf{C}], [\mathbf{I};\mathbf{C}]),\\
    \tilde{\mathbf{I}} &= \operatorname{CrossAttn}_i(\mathbf{I}, [\mathbf{U};\mathbf{C}], [\mathbf{U};\mathbf{C}]).
\end{align}

The resulting token banks are summarized by separate attentive pooling heads. For an arm $\star\in\{u,i\}$ with token matrix $\tilde{\mathbf{X}}\in\mathbb{R}^{L\times d}$,
\begin{equation}
    \mathbf{A}_{\star}
    =
    \operatorname{softmax}_{L}
    \left(
    \tilde{\mathbf{X}}W_{\star}
    \right),
    \qquad
    \mathbf{Q}_{\star}
    =
    \mathbf{A}_{\star}^{\top}\tilde{\mathbf{X}}.
\end{equation}
The two branches use independent pooling heads and may allocate different numbers of query tokens, reflecting the asymmetry between user-side and item-side information. The pooled outputs are concatenated to form the base query bank consumed by the subsequent interaction backbone.

\subsubsection{Candidate-Aware Query Bias}
\label{app:qbias}

To inject target-aware sequence evidence into query construction, we additionally use a candidate-aware query bias inspired by DIN \cite{zhou2018deep}. Given candidate representation $\mathbf{c}$ and behavior token $\mathbf{h}_i$, an interaction score is computed from
\begin{equation}
    [\mathbf{h}_i;\mathbf{c};
    \mathbf{h}_i\odot\mathbf{c};
    \mathbf{h}_i-\mathbf{c}],
\end{equation}
followed by an MLP and normalized pooling over valid sequence positions. The resulting target-aware sequence summary is used as an additive bias for the corresponding query representation. This component allows candidate information to influence the sequence path before downstream cross-domain interaction.

\subsubsection{SwiGLU Sequence Encoder}
\label{app:swiglu}

The sequence encoder uses a SwiGLU feed-forward unit from the GLU family \cite{shazeer2020glu}. Given an input projection split into two parts,
\begin{equation}
    [\mathbf{x}_1;\mathbf{x}_2]
    =
    W_{\mathrm{in}}\mathbf{x},
\end{equation}
the transformation is
\begin{equation}
    \operatorname{SwiGLU}(\mathbf{x})
    =
    W_{\mathrm{out}}
    \left(
    \mathbf{x}_1
    \odot
    \operatorname{SiLU}(\mathbf{x}_2)
    \right).
\end{equation}
The gated path modulates the carried representation and provides a stronger nonlinear sequence encoder than a standard two-layer feed-forward block.

\subsubsection{Per-Domain Sequence Tokenization and Temporal Ordering}
\label{app:domain_tokenizer}

The behavioral histories are organized into multiple domains with distinct semantics and sequence-length distributions. We therefore retain an independent tokenizer for each behavior domain. Before entering the interaction blocks, the domain-wise token sequences are concatenated while preserving their domain identities.

We additionally apply stable temporal ordering based on time-bucket information. Stable sorting preserves the relative order among events assigned to the same time bucket, while enforcing a consistent temporal organization across the concatenated sequence representation.

\subsubsection{Item-Time Query Context}
\label{app:item_time}

We further explored an item-side temporal query-context token. Item-side time or age-related dense features are projected into an additional context token, which is injected into the key/value bank used by DualQ:
\begin{equation}
    \mathbf{C}'
    =
    [\mathbf{C};\mathbf{t}_{\mathrm{item\mbox{-}time}}].
\end{equation}
This allows the generated interaction queries to condition not only on user and candidate-item representations, but also on temporal properties of the candidate item.

\subsubsection{Multi-View Sequence Modeling}
\label{app:multiview}

A single sequence truncation length may emphasize only one temporal scale of user behavior. We therefore explored multi-view sequence modeling, where each behavior domain is represented by multiple ratio-based views. In the reported configuration, one view uses the full retained sequence ($r=1.0$), while another uses a shorter recent subset ($r=0.7$).

Each view is encoded independently and participates in its own sequence interaction path. This allows the model to combine information from longer-term behavior coverage and a more recent truncated view without forcing both scales into a single sequence representation.

\subsubsection{User-Dense Tokenization Variants}
\label{app:user_dense}

Several user-dense tokenization strategies were explored during the competition. One variant separates the dense tail into a small head-of-dense subset and the remaining newly introduced features, with independent projection paths. Another constructs three tokens from fid61, fid87, and the full dense tail.

The strongest user-dense variant groups fields according to their semantic roles. Specifically, statistical fields are grouped into a token $g_{\mathrm{stat}}$, while continuous representation-like fields are grouped into $g_{\mathrm{float}}$. The two groups are independently projected before entering the subsequent interaction path. This design preserves coarse semantic distinctions among user-dense signals while keeping the token count compact.

\subsubsection{Weighted Residual Gated Pair Pooling}
\label{app:pair_pooling}

For pairwise user--item features, we explored a weighted residual gated pooling mechanism. The original pair fusion path aggregates integer embeddings, dense projections, and dense statistics. We retain this main path and add a small gated residual branch.

For dense pair fields, normalized learned weights are used to form a weighted pooled representation. For selected signed features, masked softmax weights preserve the sign structure during aggregation. Let $\mathbf{p}_{\mathrm{base}}$ denote the original pair representation and $\mathbf{p}_{\mathrm{weighted}}$ the weighted pooled representation. The residual correction is written as
\begin{equation}
    \mathbf{p}
    =
    \mathbf{p}_{\mathrm{base}}
    +
    \sigma(\mathbf{g})
    \odot
    \mathbf{p}_{\mathrm{weighted}}.
\end{equation}
This design augments the original pair fusion without replacing it, allowing informative pair fields to receive non-uniform contributions.

\subsubsection{Field-Aware Item-Dense Split}
\label{app:tokenization}

A further refinement targets item-dense field 129, which is a 130-dimensional vector. The implementation partitions the field into a 128-dimensional body vector, a scalar statistic, and a final unused coordinate:
\begin{equation}
    \mathbf{b}
    =
    \mathbf{x}_{129}[0{:}128],
    \qquad
    s
    =
    \mathbf{x}_{129}[128],
    \qquad
    u
    =
    \mathbf{x}_{129}[129].
\end{equation}
The body vector and scalar statistic are projected into separate tokens, while coordinate $u$ is not used.

The body token is computed as
\begin{equation}
    \mathbf{t}_{\mathrm{body}}
    =
    \operatorname{LN}_d
    \left(
    \operatorname{SiLU}
    \left(
    W_b
    \operatorname{LN}_{128}(\mathbf{b})
    +
    \mathbf{c}_b
    \right)
    \right),
\end{equation}
while the scalar is first compressed:
\begin{equation}
    \tilde{s}
    =
    \log(1+\max(s,0)),
\end{equation}
and then projected as
\begin{equation}
    \mathbf{t}_{\mathrm{stat}}
    =
    \operatorname{LN}_d
    \left(
    \operatorname{SiLU}
    (W_s\tilde{s}+\mathbf{c}_s)
    \right).
\end{equation}

The two slices differ sharply in dimensionality and preprocessing. A single projection can allow the 128-dimensional body to dominate the one-dimensional statistic. Splitting the field before projection gives both components independent token-level representations.

\subsubsection{Additional Temporal Context}
\label{app:temporal}

Beyond the item-time query token, the competition system uses several complementary temporal signals.

Relative sequence order is encoded with rotary position embeddings (RoPE) \cite{su2021roformer}. For a rotary subspace with inverse frequencies $\omega_k$, the position-dependent transformation is
\begin{equation}
    \operatorname{RoPE}(\mathbf{x},m)
    =
    \mathbf{x}\odot\cos\boldsymbol{\theta}_m
    +
    \operatorname{rothalf}(\mathbf{x})
    \odot
    \sin\boldsymbol{\theta}_m.
\end{equation}

A global-time token is constructed from absolute and cyclic calendar features:
\begin{equation}
    \mathbf{t}_{\mathrm{time}}
    =
    \operatorname{SiLU}
    \left(
    W_g
    [\mathbf{e}_1;\ldots;\mathbf{e}_F]
    \right).
\end{equation}

Within each behavior domain, inter-event gaps are bucketized and mapped to embeddings. A domain-specific gate controls their contribution:
\begin{equation}
    \mathbf{h}_i^{(\ell)}
    \leftarrow
    \mathbf{h}_i^{(\ell)}
    +
    \boldsymbol{\gamma}^{(\ell)}
    \odot
    \mathbf{g}_i.
\end{equation}

We also compute a recency-weighted sequence summary. Let
$\delta_i=\log(1+\Delta_i^{\mathrm{days}})$ denote the log time since event and $g_i$ the inter-event gap. The recency logit is
\begin{equation}
    \lambda_i
    =
    -\alpha\delta_i-\beta g_i,
\end{equation}
with normalized weight
\begin{equation}
    w_i
    =
    \frac{\exp(\lambda_i)}
    {\sum_{j\in\mathcal{V}^{(\ell)}}\exp(\lambda_j)}.
\end{equation}
The resulting time-decay summary is
\begin{equation}
    \mathbf{u}^{(\ell)}
    =
    \sum_{i\in\mathcal{V}^{(\ell)}}
    w_i
    \tilde{\mathbf{h}}_i^{(\ell)}.
\end{equation}

\subsubsection{Optimization Stabilization}
\label{app:optimization}

During training, an exponential moving average of dense model weights is maintained:
\begin{equation}
    \bar{\boldsymbol{\theta}}_t
    =
    \rho
    \bar{\boldsymbol{\theta}}_{t-1}
    +
    (1-\rho)
    \boldsymbol{\theta}_t.
\end{equation}
Validation and inference use the averaged weights $\bar{\boldsymbol{\theta}}$, following common weight-averaging practices \cite{izmailov2018swa,tarvainen2017mean}.

The training objective supports binary cross-entropy and focal loss under class imbalance \cite{lin2017focal}. For prediction probability $p$ and label $y\in\{0,1\}$,
\begin{equation}
    \mathcal{L}_{\mathrm{focal}}(p,y)
    =
    -\alpha_t
    (1-p_t)^\gamma
    \log p_t,
\end{equation}
where
\begin{equation}
    p_t
    =
    yp+(1-y)(1-p).
\end{equation}

\subsection{Experimental Details}
\label{app:experiments}

\subsubsection{Experimental Setup and Reproducibility}

The model is evaluated on the Tencent Advertising Algorithm Competition dataset. The training split contains approximately 19.78 million samples, and the validation split contains approximately 2.21 million samples. The positive conversion rate is close to eight percent, and AUC is used as the primary evaluation metric.

The competition experiments were conducted on the official platform. The reported configurations use multi-GPU distributed training when several GPUs are available, bfloat16 automatic mixed precision, and checkpoint sidecars that preserve the schema and feature wiring required for inference. Because the competition-stage experiments span several successive configurations, hyperparameters may differ across checkpoints; the corresponding results should therefore be interpreted as a record of system development rather than a single strictly controlled leave-one-out ablation.

\subsubsection{Relation to the Official Baseline}

The official competition baseline uses a HyFormer-style query-based backbone with RankMixer non-sequential tokenization, a SwiGLU sequence encoder with RoPE, and query-based sequence fusion \cite{huang2026hyformer}. Our competition system extends this starting point with additional feature tokenization, query construction, temporal modeling, long-tail representation, and optimization components.

\begin{table*}[t]
\centering
\caption{Configuration and module comparison between the official competition baseline and our competition configuration.}
\label{tab:app_baseline}
\small
\begin{tabular}{lcc}
\toprule
 & Official baseline & Our configuration \\
\midrule
Model dimension $d_{\mathrm{model}}$ & 80 & 192 \\
Query tokens & 2 & 6 \\
Interaction blocks & 2 & 2 \\
User / item NS tokens & 4 / 2 & 14 / 3 \\
\midrule
RankMixer NS tokenization & Yes & Yes \\
SwiGLU encoder with RoPE & Yes & Yes \\
DualQ query generation & -- & Yes \\
Candidate-aware query bias & -- & Yes \\
Field-aware dense tokenization & -- & Yes \\
Global-time token & -- & Yes \\
Sequence time-gap buckets / domain gates & -- & Yes \\
Time-decay behavior summary & -- & Yes \\
ID-frequency buckets & -- & Yes \\
fid-87 residual token & -- & Yes \\
Exponential moving average & -- & Yes \\
\bottomrule
\end{tabular}
\end{table*}

\subsubsection{Competition-Stage Results}

Table~\ref{tab:app_progression} summarizes the principal competition-stage experiments. Results are ordered by Test AUC for readability. These rows should not be interpreted as a strict cumulative ablation chain: several experiments correspond to parallel architectural variants evaluated at different stages of system development.

\begin{table*}[t]
\centering
\caption{Competition-stage results ordered by Test AUC. Some rows are parallel architectural variants rather than strictly cumulative additions.}
\label{tab:app_progression}
\small
\begin{tabular}{lcc}
\toprule
Configuration / Idea & Val AUC & Test AUC \\
\midrule
Round-2 baseline + preliminary-stage transfers
    & 0.839385 & 0.830408 \\
ID-frequency buckets
    & 0.839263 & 0.832477 \\
Capacity scaling ($d$: 128$\rightarrow$192, queries: 4$\rightarrow$6)
    & 0.839664 & 0.832935 \\
Item-dense field-wise tokenizer
    & 0.840889 & 0.834219 \\
SwiGLU sequence encoder
    & 0.841126 & 0.834723 \\
Per-domain tokenizer + stable temporal sort
    & 0.841516 & 0.834823 \\
Item-time query-context token
    & 0.841619 & 0.834913 \\
Multi-view sequence modeling
    & 0.841571 & 0.834973 \\
User-dense tail with separate projections
    & 0.841774 & 0.835186 \\
User-dense three-token design
    & 0.841773 & 0.835219 \\
User-dense tail, untokenized
    & 0.841770 & 0.835220 \\
Two-group semantic user-dense tokens
    & 0.841996 & 0.835393 \\
Weighted residual gated pair pooling
    & 0.842041 & 0.835412 \\
Item-dense field-129 two-token split
    & \textbf{0.842206} & \textbf{0.835594} \\
\bottomrule
\end{tabular}
\end{table*}

Several trends can be observed from the competition-stage results. First, field-aware handling of dense inputs consistently becomes more important as the system matures. Moving from a coarse item-dense representation to field-wise tokenization produces a substantial improvement, while the final field-129 split further yields the best reported result.

Second, the user-dense experiments show that finer tokenization is not automatically beneficial. Separate projections and a three-token design perform similarly to the untokenized tail, whereas grouping the fields by semantic role into statistical and continuous groups provides a clearer improvement. This suggests that token boundaries should follow meaningful feature structure rather than simply increasing token count.

Third, temporal and multi-view sequence variants provide smaller but consistent gains. The item-time query-context token and ratio-based multi-view sequence modeling improve the test score beyond the preceding sequence configuration, indicating that candidate-side temporal context and multiple sequence horizons can complement the main interaction path.

Finally, the weighted residual gated pair pooling and field-129 split provide the strongest late-stage improvements. Both designs preserve the original representation path while adding structure-aware auxiliary routes, suggesting that selective refinement of heterogeneous feature groups can be more effective than indiscriminate expansion of model capacity.

\end{document}